\documentclass[11pt]{iopart}

\usepackage{graphicx}
\usepackage[utf8]{inputenc}
\usepackage[T1]{fontenc}
\usepackage{mathptmx}
\usepackage{mwe}
\usepackage{subfigure}
\usepackage{multirow}
\usepackage{makecell}
\begin{document}

\title[Neutron hardness of EJ-276 scintillation material]{Neutron hardness of EJ-276 scintillation material}

\author{S.~Mianowski$^1$, K.~Brylew$^1$ , A.~Dziedzic$^1$, K.~Grzenda$^1$, P.~Karpowicz$^1$, A.~Korgul$^2$, M.~Krakowiak$^2$, R.~Prokopowicz$^1$, G.~Madejowski$^1$, Z.~Mianowska$^1$, M.~Moszynski$^1$, T.~Szczesniak$^1$ and M.~Ziemba$^1$}

\address{$^1$National Centre for Nuclear Research, Otwock, A. Soltana 7, Poland
}
\address{$^2$Faculty of Physics, University of Warsaw,  Warsaw, Pasteura  5, Poland
}

\ead{s.mianowski@ncbj.gov.pl}

\date{\today}

\begin{abstract}

This paper presents the results of the fast neutron irradiation (E$_n$ > 0.5~MeV) of an EJ-276 scintillator performed in the MARIA research reactor with fluence up to 5.3$\times$10$^{15}$ particles/cm$^2$. In our work, four samples  with size $\phi$25.4~mm$\times$5~mm were tested. The changes in the light yield, emission and absorption spectrum and neutron/gamma discrimination using PuBe source before and after irradiation are presented. The figure of merit in neutron/gamma discrimination based on the charge integration method for different neutron fluences and different short gate integration times are determined.

\end{abstract}

\maketitle

\section{Introduction}
In the past, an effort to develop efficient and low-cost fast neutron detectors for nuclear science, high energy physics and homeland security has grown significantly. It  was  shown  by Brooks \cite{Brooks1959, Brooks1960} that  plastic  scintillators can also present discrimination capabilities. However, the results in neutron/gamma (n/$\gamma$) discrimination were worse in comparison to a liquid organic scintillators \cite{Winyard, Hansen, Woolf, Iwanowska}. On the other hand, toxicity and flammability of liquid scintillators eliminate them from a wide range of use. The situation with plastic scintillator with pulse shape discrimination capabilities changed about a decade ago, where a new plastic EJ-299-33 and EJ-299-34 from Eljen Technology \cite{EJ} appears and their properties were explored extensively \cite{Iwanowska, Zaitseva2018}.

Presented in this paper, the EJ-276 scintillator from Eljen Technology replaces all versions of EJ-299-33 and EJ-299-34 PSD scintillators \cite{EJ, Zaitseva2018}. The response mechanism to neutron and gamma radiation was described in \cite{Zaitseva2012}. In general, a more extended time response to neutrons in comparison to gamma quanta allows us to construct Pulse Shape Discrimination (PSD) algorithms to detect the neutrons in the presence of gamma radiation.

The goal of this work is to study EJ-276 radiation hardness by the change of scintillator properties induced by fast neutrons. Characteristics like light emission spectra, absorption spectra, light yield and n/$\gamma$ discrimination capabilities are shown. For this reason four samples of EJ-276 scintillators were tested and three of them were irradiated by high neutron flux available at the National Centre for Nuclear Research (NCBJ). 

\section{Experimental set-up and data analysis}

Four samples of EJ-276 scintillator with sizes $\phi$~25.4~mm~$\times$~5~mm were used in our measurements. Table \ref{tab:ej276} presents the basic properties of the chosen scintillator \cite{EJ}. Two previous generation plastic scintillators: EJ-299-33 and EJ-299-34 are also compared. The EJ-299-34 do not have the official datasheet and the light yield reference is not available, but the photo-electron number comparison can be found in literature \cite{Iwanowska}. A general characterisation of an EJ-276 scintillator for non-irradiated samples was also presented in \cite{Zaitseva2018, Martyna}.

\begin{table}[!h]
\caption{EJ-276 scintillator basic properties \cite{EJ}. EJ-299-33 and EJ-299-34 plastic scintillators are presented for comparison \cite{Iwanowska}.}
\label{tab:ej276}
\centering
\begin{tabular}{c|c|c|c|c}
\hline
\multicolumn{2}{c|}{Parameter} 									& 		EJ-276		& EJ-299-33   & EJ-299-34  \\ \hline
\multicolumn{2}{c|}{Light yield (photons/MeV e$^-$)}  			&		8600     	& 8600  & -	   \\
\multicolumn{2}{c|}{Density (g/cm$^3$)}					   		& 		1.096	    & 1.08  & - \\
\multirow{2}{*}{\makecell{Mean decay time \\of three components (ns)}} & Gamma excitation	& 	13, 35, 270	& 4.6, 19, 130 & 4.3, 18, 140 \\ 
  												   & Neutron excitation &	13, 59, 460	& 5, 22, 180 &	4.5, 20, 170 	\\ \hline
\end{tabular}
\end{table}

\subsection{Irradiation process}
\begin{figure}[!htb]
  \centering
  \def\svgwidth{200pt}
  \includegraphics[width=0.75\textwidth]{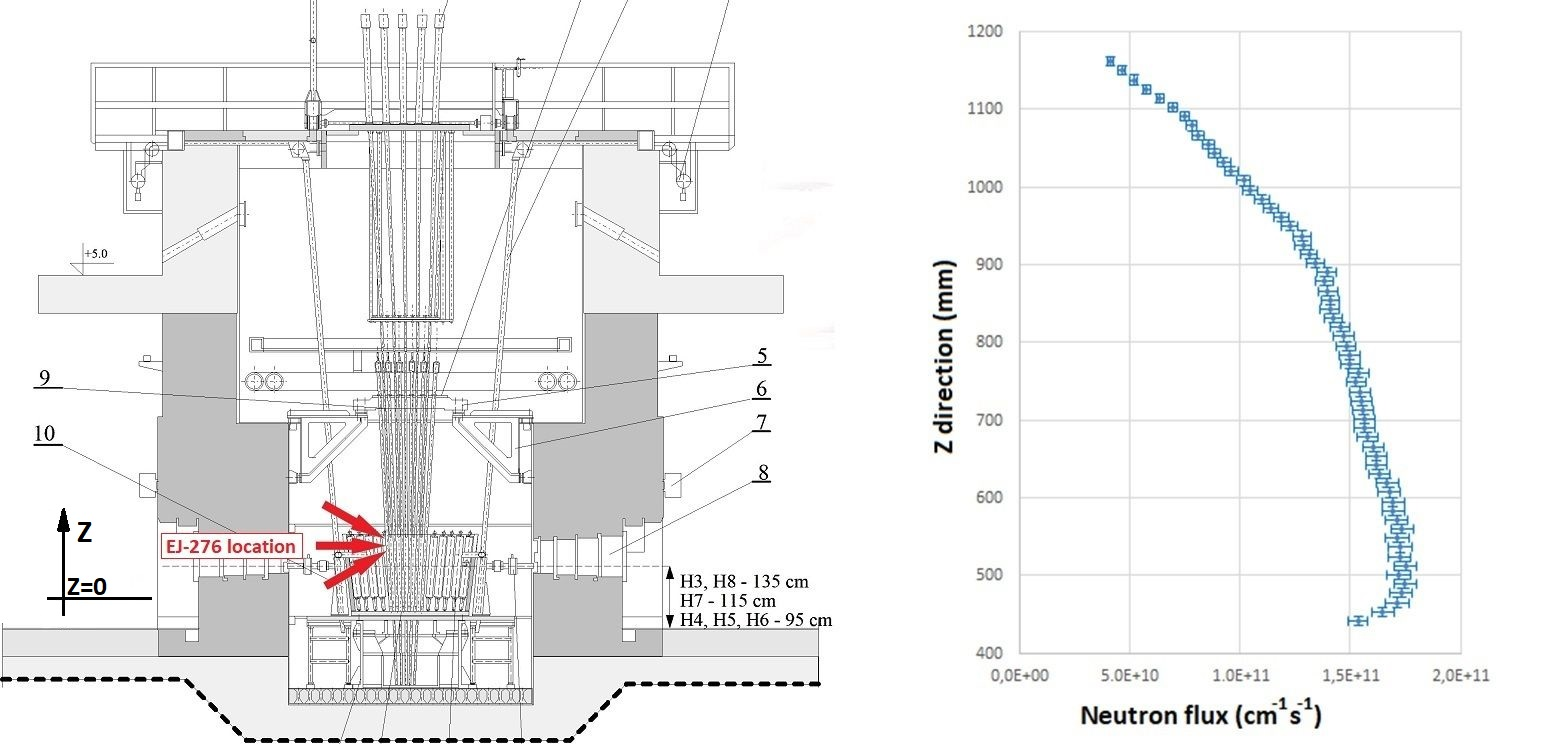}
  \caption{ EJ-276 samples location in the MARIA research reactor during irradiation session (left) and a fast neutron flux distribution in the Z direction (right).}
  \label{fig:maria}
\end{figure}

The irradiation sessions were done at the MARIA research reactor at NCBJ. The MARIA is a high neutron flux (see Figure \ref{fig:maria}) research reactor with 30~MW of nominal thermal power and fuel channels in a conical matrix of beryllium blocks surrounded by a graphite reflector. 

During the irradiation process, each sample (three out of four) was exposed to different neutron flux distribution. The neutron flux differentiation in the Z direction (orthogonal to the earth plane) in the reactor was exploited in this case (see Figure \ref{fig:maria}). In combination with different irradiation times, it gives us the opportunity for better neutron fluence differentiation. The magnitude change in the order of 1000 in neutron fluence was achieved in this case. Table \ref{tab:irrtime} shows irradiation time, neutron flux and the fluence obtained for a single EJ-276 sample. 

\begin{table}[!t]
\caption{EJ-276 scintillator irradiation time and neutron flux, which correspond to the fast neutron fluence obtained. Neutron flux and neutron fluence are known with 3$\%$ uncertainty.}
\label{tab:irrtime}
\centering
\begin{tabular}{c|c|c|c}
\hline
EJ-276 		& 		Irradiation 		& 	Neutron flux	  		&	Neutron fluence	  \\ 
sample  	& 		time (s)      		& 		(cm$^{-2}$ s$^{-1}$)	  & 	(cm$^{-2}$)    	 \\ \hline
1.  			& 			17     		& 		0.46$\times$10$^{12}$ &		7.8$\times$10$^{12}$   \\
2.   		& 			460       		& 		1.06$\times$10$^{12}$ &		4.9$\times$10$^{14}$   \\
3.    		& 			3060 			& 		1.74$\times$10$^{12}$ &		5.3$\times$10$^{15}$  \\ \hline
  
\end{tabular}
\end{table}

In the next step, all samples were stored for period of one month in a safety room to reduce their internal activity from neutron activation. Gamma spectroscopy analysis made with a germanium detector for the third sample showed, that  major components of EJ-276 scintillator did not exhibit their internal activity after this period time. 

\begin{figure}[!htb]
  \centering
  \def\svgwidth{200pt}
  \includegraphics[width=0.65\textwidth]{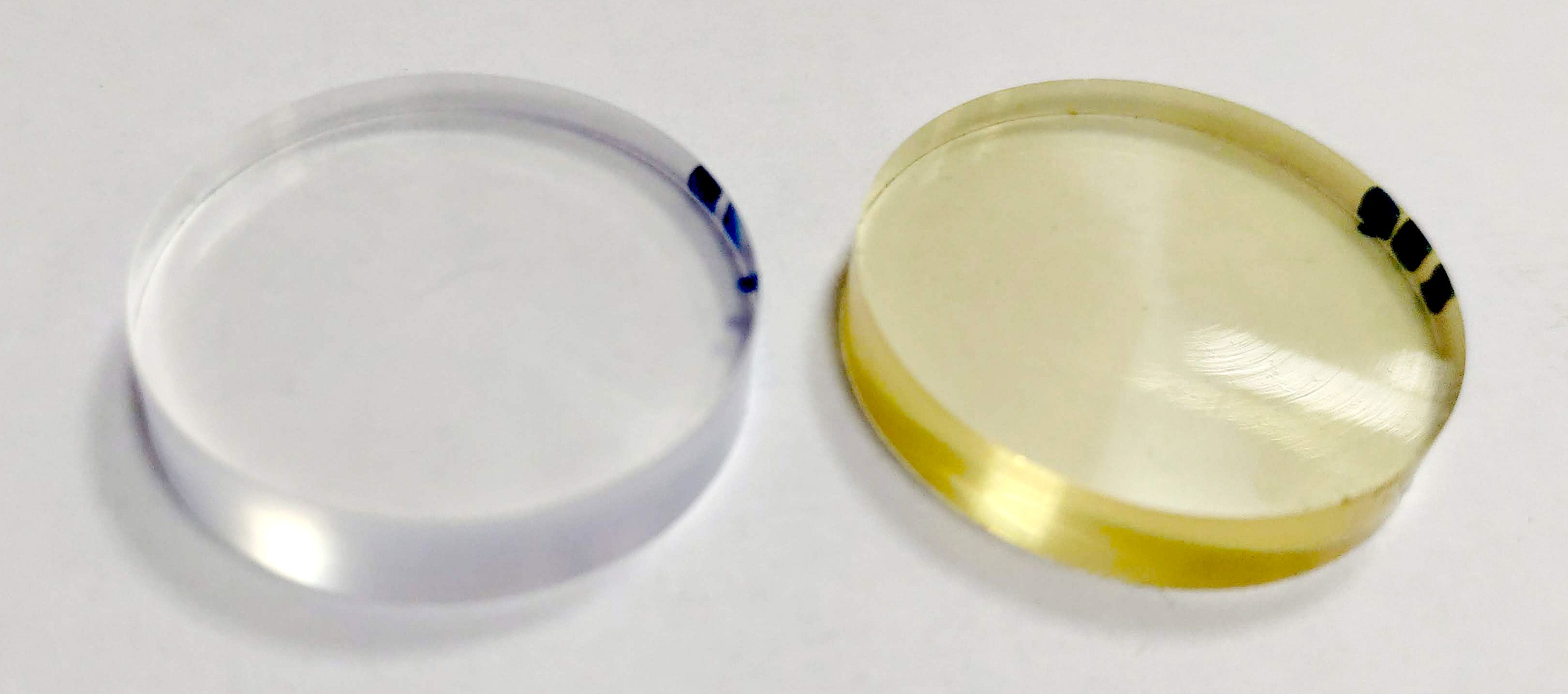}
  \caption{ Two EJ-276 samples. Before irradiation -- left and after the longest irradiation time in the MARIA research reactor -- right.}
  \label{fig:foto}
\end{figure}

Figure \ref{fig:foto} shows two EJ-276 scintillator samples before and after the longest irradiation. The colour change is clearly seen in the second case.

\subsection{EJ-276 radioluminescence emission and absorption spectrum}

After irradiation, the X-ray excited radioluminescence (RL) emission spectrum and absorption spectrum by registering sample transmission were measured.

\begin{figure}[!htb]
  \centering
  \def\svgwidth{200pt}
  \includegraphics[width=0.5\textwidth]{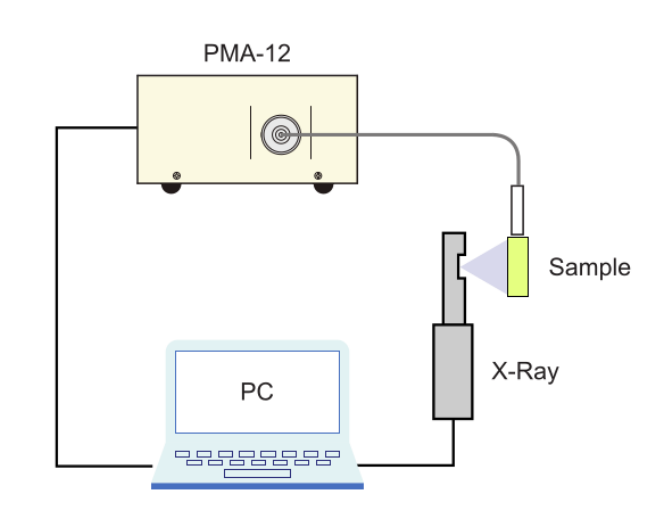}
  \caption{Experimental set-up for EJ-276 radioluminescence spectrum registration.}
  \label{fig:PMA_Xray}
\end{figure}

\begin{figure}[!htb]
  \centering
  \def\svgwidth{200pt}
  \includegraphics[width=1\textwidth]{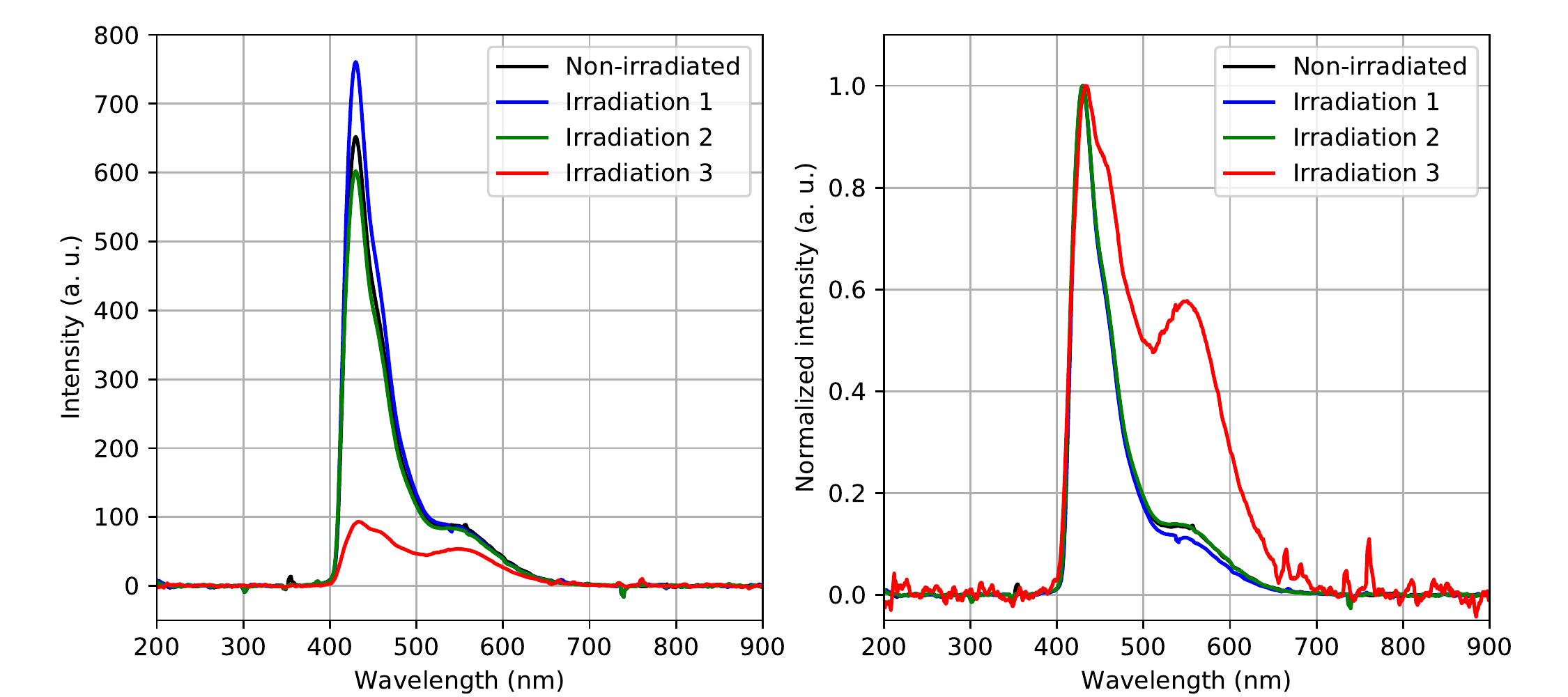}
  \caption{Experimental radioluminescence spectra registered for different EJ-276 scintillator  irradiation times. Left -- spectra for the same time of data collection. Right -- spectra normalised to the same amplitude.}
  \label{fig:emis}
 \end{figure}

The RL excitation was performed using a tungsten X-ray tube operating at 100~$\mu$A and 10~kV. The luminescence spectrum was registered using a Hamamatsu Photonic Multichannel Analyser (PMA) C10027-01 measured in the range of 200~to~900~nm. The PMA has a resolution of <2~nm.  The integration time was 15~s and the spectra have been averaged after 10 repetitions. Dark current spectrum was subtracted before each measurement and calibration curves provided by the manufacturer were applied. The experimental set-up is presented in Figure \ref{fig:PMA_Xray}. The results obtained are presented in Figure \ref{fig:emis}.

As can be seen, there is no significant difference in the shape spectrum for neutron fluence up to 4.9$\times$10$^{14}$. The significant light reduction is visible for the highest neutron fluence, especially in the range of 440~nm. Differences in the light amount are discussed in the following section.

\begin{figure}[!htb]
  \centering
  \def\svgwidth{200pt}
  \includegraphics[width=0.5\textwidth]{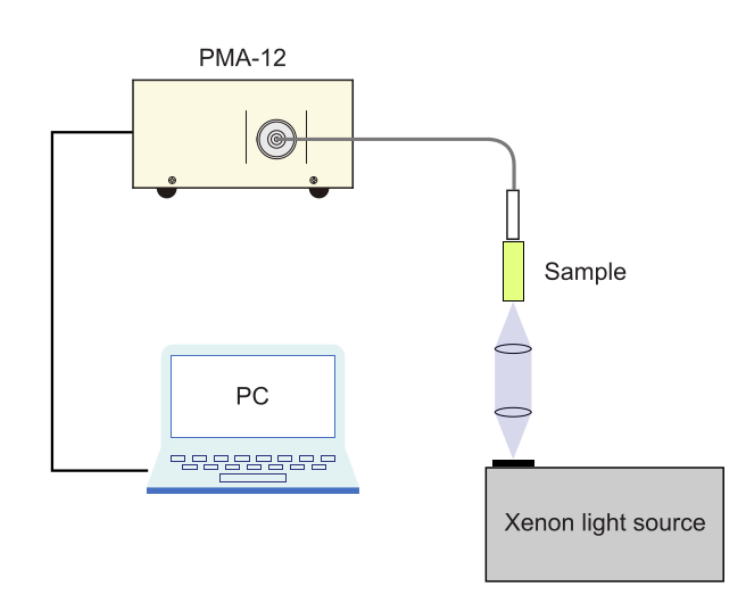}
  \caption{Experimental set-up for EJ-276 absorption spectrum registration.}
  \label{fig:PMA_abs}
\end{figure}

\begin{figure}[!htb]
  \centering
  \def\svgwidth{200pt}
  \includegraphics[width=0.7\textwidth]{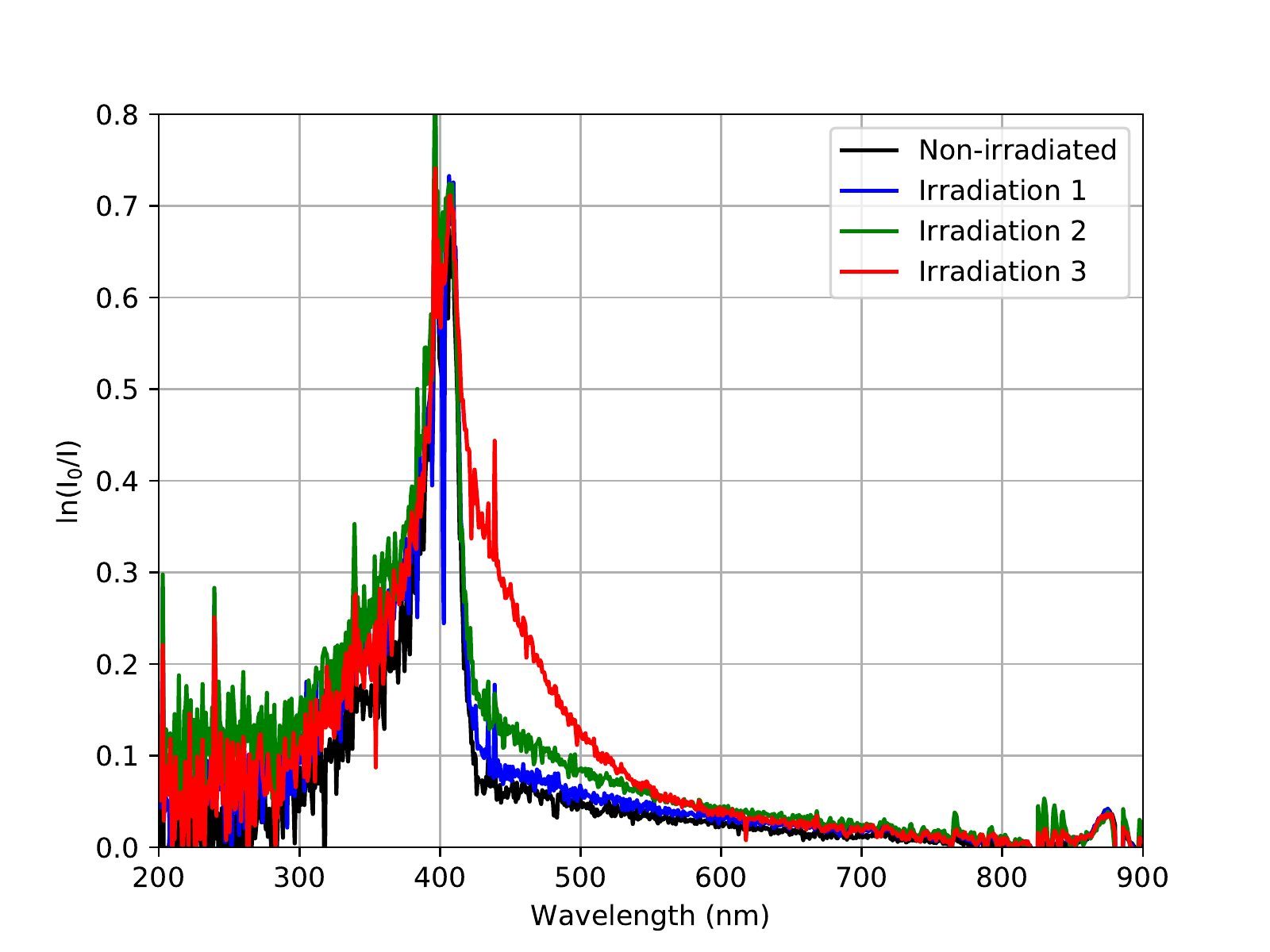}
  \caption{Experimental absorption spectra registered for different EJ-276 scintillator  irradiation times. Spectra were normalised to the same amplitude.}
  \label{fig:abs}
 \end{figure}

PMA-12 has also been used to measure the absorption spectrum. A Spectral Products ASB-XE-175EX xenon lamp was used as a source of white light. In the first step the spectrum of the xenon lamp was measured unobstructed by the investigated sample (Figure \ref{fig:PMA_abs}). Then, each sample was placed on the path of the light beam and the spectrum was recorded once again. The wavelength-dependent absorbance (A) was calculated using the formula:
\begin{equation}
\textnormal{A}(\lambda) = -\textnormal{ln} \frac{\textnormal{I}(\lambda)}{\textnormal{I}_0(\lambda)},
\end{equation}
where I$_0$ is the initial beam intensity at given wavelength $\lambda$ and I is the intensity of the beam after passing through the sample. The dark count spectrum was subtracted, and similarly, the calibration curves have been applied. The integration time was 30~s and the measurements were repeated 30~times and then averaged. Once again, there is no significant difference in the spectrum obtained for neutron fluence up to 4.9$\times$10$^{14}$~cm$^{-2}$ (Figure \ref{fig:abs}). The significant absorbance increase is clearly visible for the highest neutron fluence in the region over 420~nm. This spectrum part strongly overlaps with the emission spectrum and, as a result, the light emission reduction is observed.

\subsection{Light yield change}

In this part, each scintillator was wrapped in Teflon tape and coupled to the XP5500 photomultiplier (PMT) from Photonis. The PMT was supplied by the bias voltage of -1100~V. To extract the single photo-electron PMT response, the direct signal from the anode was sent to the preamplifier and, in the next step, to the ORTEC 672 spectroscopy amplifier. Finally, all data were analysed by the TUKAN multichannel analyser and saved to a PC.

\begin{table}[!t]
\caption{Calibration gamma sources used in the experiment. Full energy peaks and Compton edges are listed.}
\label{tab:calib}
\centering
\begin{tabular}{c|c|c}
\hline
Source 		& 		Full energy peak	& 	Compton edge energy	  		 \\ 
sample  	& 		(keV)	      		& 		(keV)	 \\ \hline
$^{241}$Am  		& 			59.5    & 		(not used)  \\
$^{137}$Cs   		& 			661.7   & 		477.3  \\
$^{54}$Mn    		& 			834.8		& 		639.2  \\
$^{22}$Na    		& 			511.0 	& 		340.7  \\ 
$^{22}$Na    		& 			1274 	& 		1061.7  \\ \hline
  
\end{tabular}
\end{table}

\begin{figure}[!h]
  \centering
  \def\svgwidth{200pt}
  \includegraphics[width=0.9\textwidth]{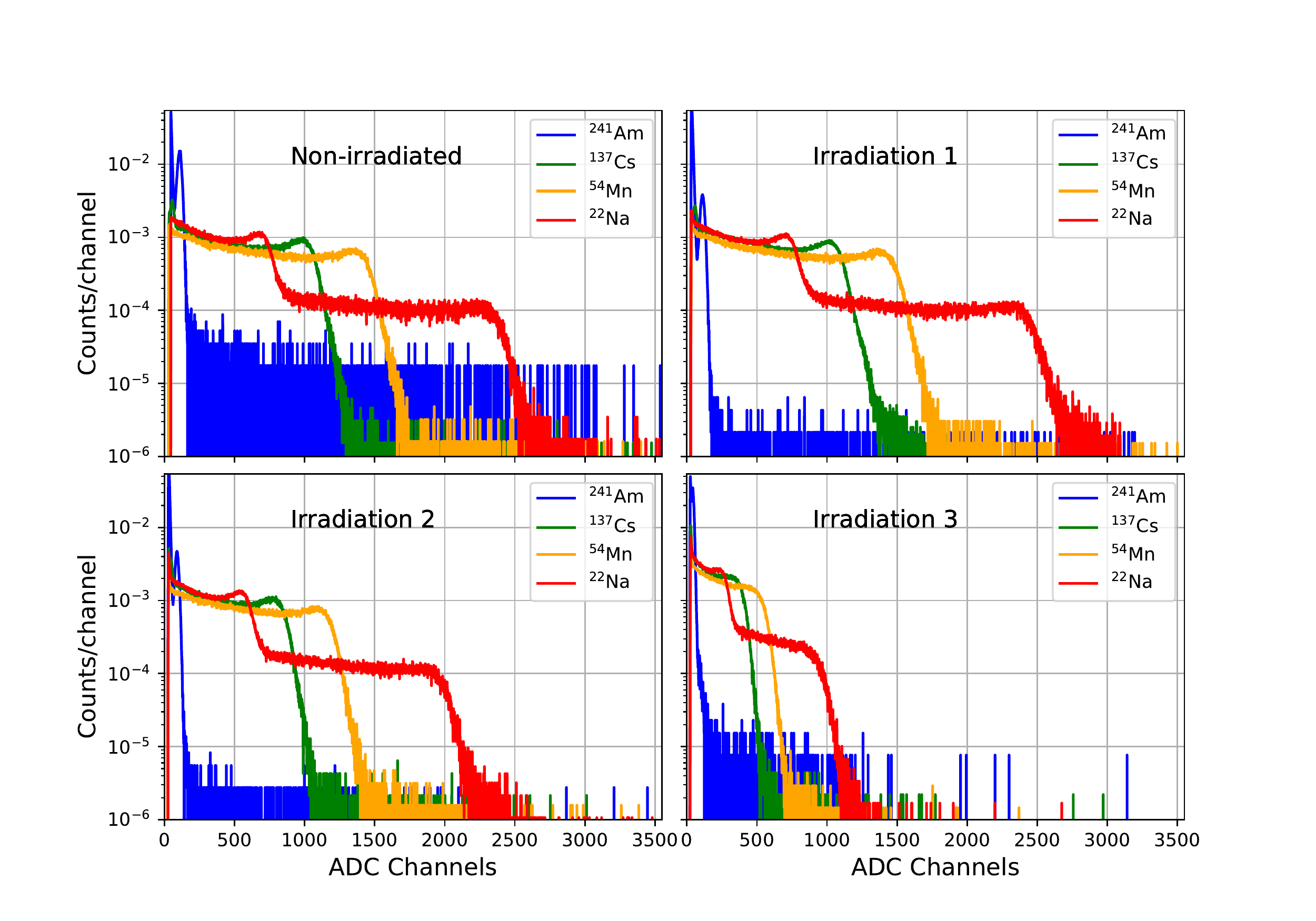}
  \caption{EJ-276 energy calibration spectra measured for different irradiation. Gamma sources like $^{241}$Am, $^{137}$Cs, $^{54}$Mn and $^{22}$Na were used in this case. All spectra were normalised to unity area. }
  \label{fig:calib_spec}
 \end{figure}

\begin{figure}[!h]
  \centering
  \def\svgwidth{200pt}
  \includegraphics[width=1\textwidth]{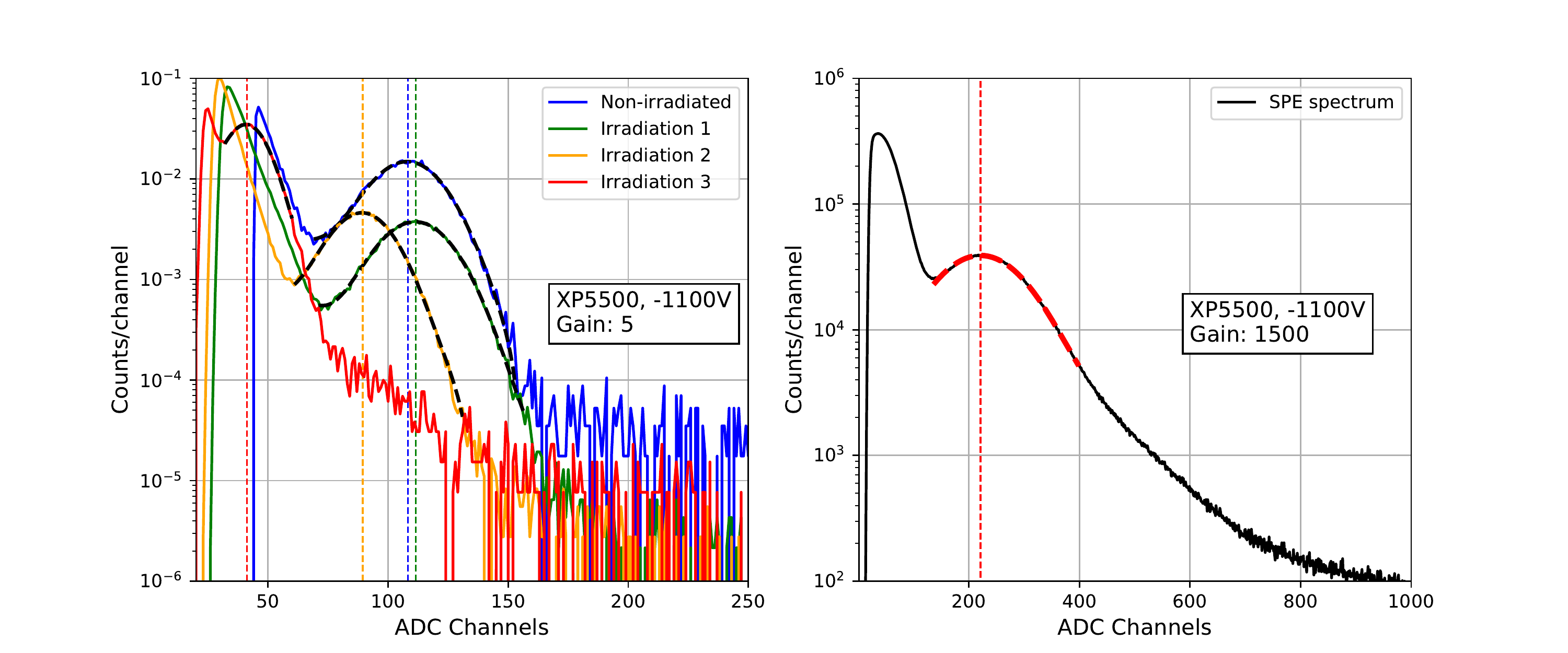}
  \caption{Comparison of the full energy peak position in $^{241}$Am for different neutron fluence (left) and the position of the single photo-electron peak in the PMT dark count spectrum. }
  \label{fig:Am241_spec}
 \end{figure}

To perform the EJ-276 energy calibration in electron equivalent scale, additional calibration gamma sources: $^{241}$Am (59.5~keV), $^{137}$Cs (661.7~keV), $^{54}$Mn (835~keV) and $^{22}$Na (511~keV, 1274~keV) were used. In brackets the full energy peaks are listed and only for $^{241}$Am was the full energy peak observed. In other cases, the Compton edges were used as reference points with an energy determination criterion of 80$\%$ of the Compton edge maximum \cite{Martyna, Lukasz}. Table \ref{tab:calib} presents listed gamma full energy peaks and corresponding Compton edge energies used in the calibration process. Figure \ref{fig:calib_spec} shows all gamma spectra registered for a different irradiation time. As can be seen, the general trend of calibration points follows the direction of the lower channel number with neutron fluence increase. This effect is seen, if the position of full energy peak in a $^{241}$Am is investigated (Figure \ref{fig:Am241_spec} left). This observation is connected with light yield reduction. The small deviation from this tendency is observed for the second irradiation.

\begin{figure}[!htb]
  \centering
  \def\svgwidth{200pt}
  \includegraphics[width=1\textwidth]{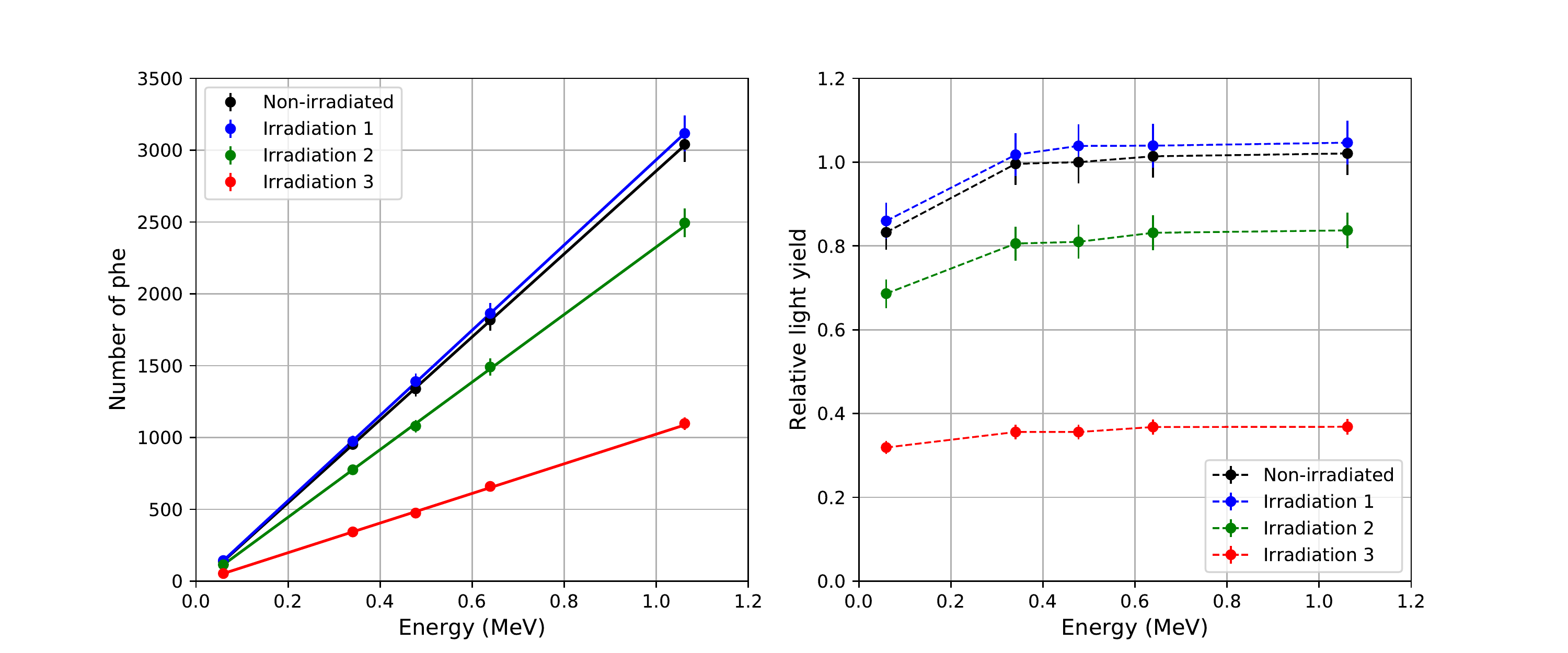}
  \caption{Left -- the photo-electron number in EJ-276 scintillator as a function of energy for different irradiation times. Lines show linear fits. Right -- relative light yield for different irradiation times. Light response to energy of 477.3~keV (Compton edge of 662~keV gamma line from $^{137}$Cs) was used as a normalisation point.}
  \label{fig:light_out}
 \end{figure}

To obtain a numerical value of the light yield before and after irradiation, the position of single photo-electron (SPE) peak in the spectrum must be determined (Figure \ref{fig:Am241_spec} right) and compared to the chosen calibration points. The average value of (2730$\pm$220) photo-electrons/MeV (phe/MeV) for non-irradiated scintillator is in an excellent agreement with values obtained for other PMTs, measured in \cite{Martyna}. For other cases, in the order of neutron fluence increases, the values are as follows: (2810$\pm$220)~phe/MeV, (2230$\pm$180)~phe/MeV and (990$\pm$60)~phe/MeV. Finally, Figure \ref{fig:light_out} shows the photo-electron number for different energies and relative light yield for different energies and different irradiation times normalised to the energy of 477.3~keV for a non-irradiated sample.

One observation is that the light yield per MeV is proportional in the explored energy range. The results are in agreement with those obtained for other plastic scintillators \cite{Nassalski, Swiderski}.

Secondly, we observe a small light yield increase (2.9$\%$ on average) after the first irradiation. This effect is also visible in the radioluminescence emission spectrum (Figure \ref{fig:emis}), where a different photodetector set-up was used and finally, in the figure of merit (FoM) calculation (see next section). Nevertheless, the characteristics observed are in agreement in the range of uncertainties  obtained, and the differences can be explained by the different sample qualities.

The last conclusion is that the change of relative light yield is energy independent in the first approximation and is in the range of 2.9$\%$, -18.3$\%$, -63.7$\%$ for subsequent irradiation times (minus indicates the light yield reduction).

\subsection{EJ-276 pulse shape discrimination}

For PSD analysis the PuBe source with activity about 28~GBq was used. This source is characterised by a continuous neutron energy spectrum up to 11~MeV and average neutron energy about $<$E$_n>$=4.5~MeV. PuBe is also a gamma emitter with lines up to 4.4~MeV. 

\begin{figure}[!htb]
  \centering
  \def\svgwidth{200pt}
  \includegraphics[width=0.4\textwidth]{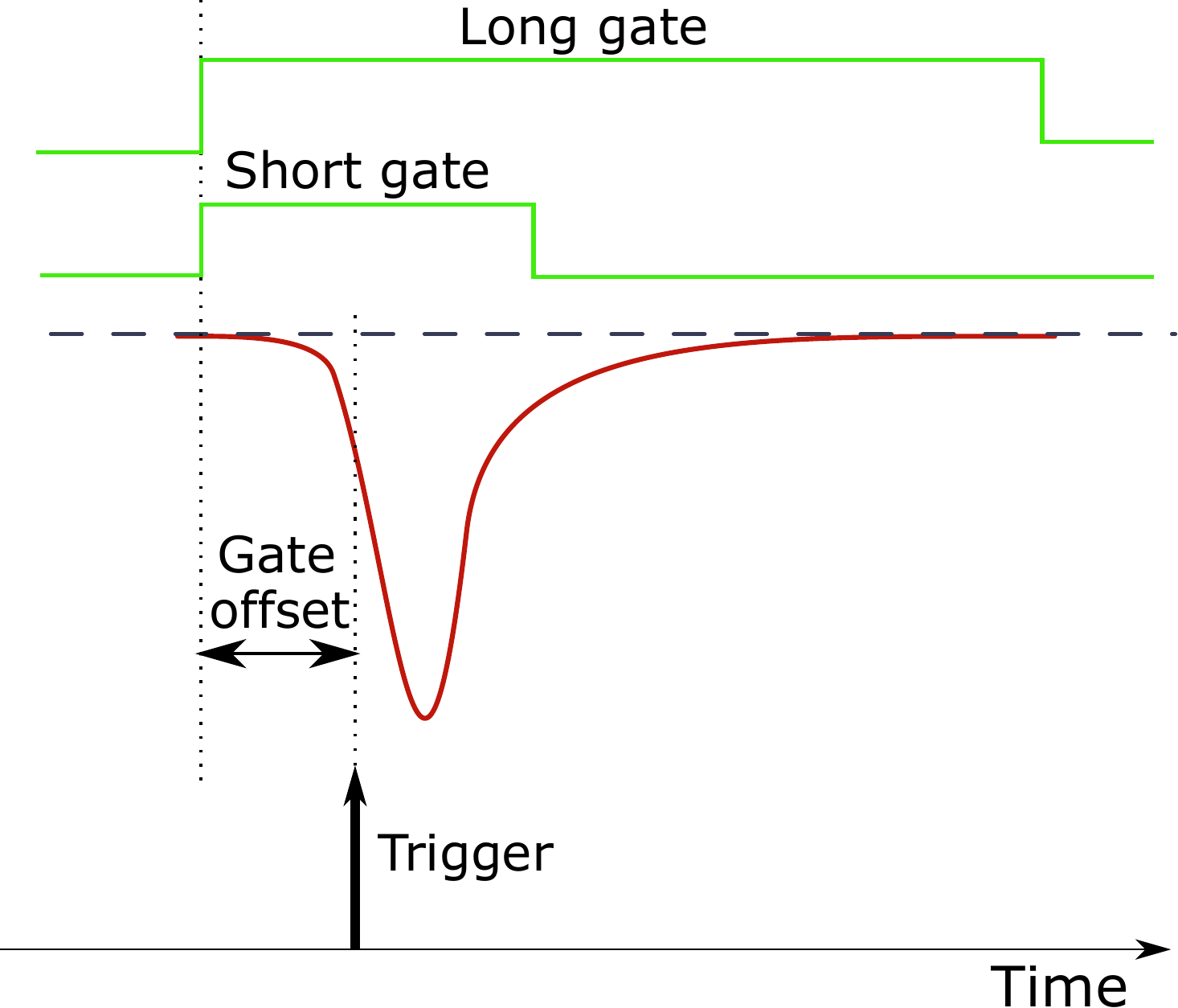}
  \caption{Illustration of the n/$\gamma$ discrimination method used in our acquisition system. The gate offset of 50 ns contributes to the length of each gate.}
  \label{fig:pulse}
\end{figure}

In our case pulse shape discrimination criterion was chosen as a ratio of a short to a long integration gate ($Q_{short}/Q_{long}$) implemented in a CAEN DT5730 digitizer \cite{CAEN} and calculated for each pulse in chosen energy range (E$\pm$10$\%$E): 100~keVee, 300~keVee, 500~keVee, 700~keVee and 1000~keVee. To perform energy calibration in an electron equivalent scale, the same gamma sources were used as before.

\begin{figure}[!h]
  \centering
  \def\svgwidth{200pt}
  \includegraphics[width=1.2\textwidth]{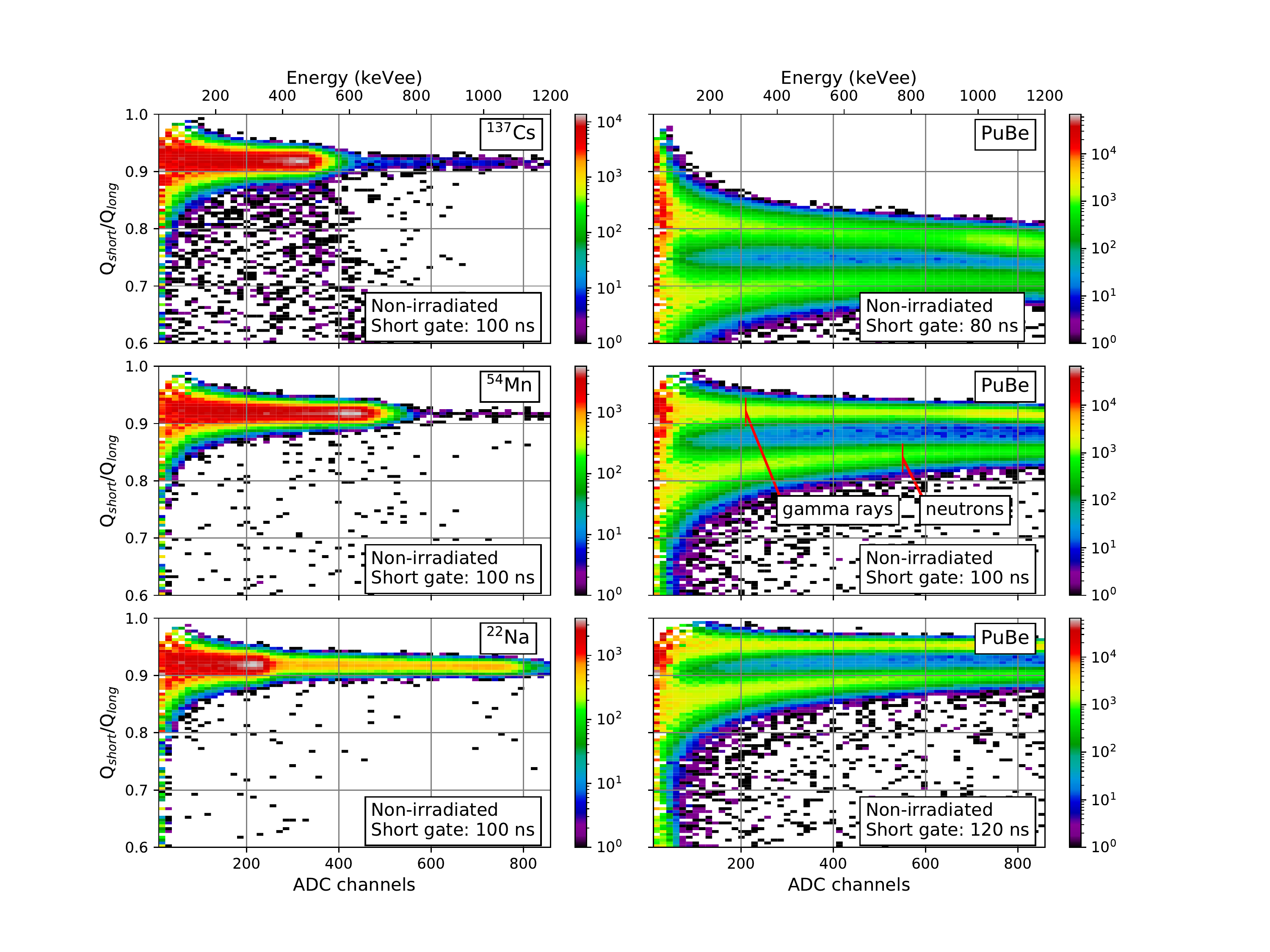}
  \caption{Examples of n/$\gamma$ discrimination spectra for non-irradiated scintillator. Left -- calibration gamma sources: $^{137}$Cs (top), $^{54}$Mn (middle) and $^{22}$Na (bottom) and short integration gate of 100~ns. Right -- PuBe spectrum obtained for short integration gate of 80~ns (top), 100~ns (middle), 120~ns (bottom).}
  \label{fig:PSD_Irr0}
 \end{figure}

\begin{figure}[!h]
  \centering
  \def\svgwidth{200pt}
  \includegraphics[width=0.8\textwidth]{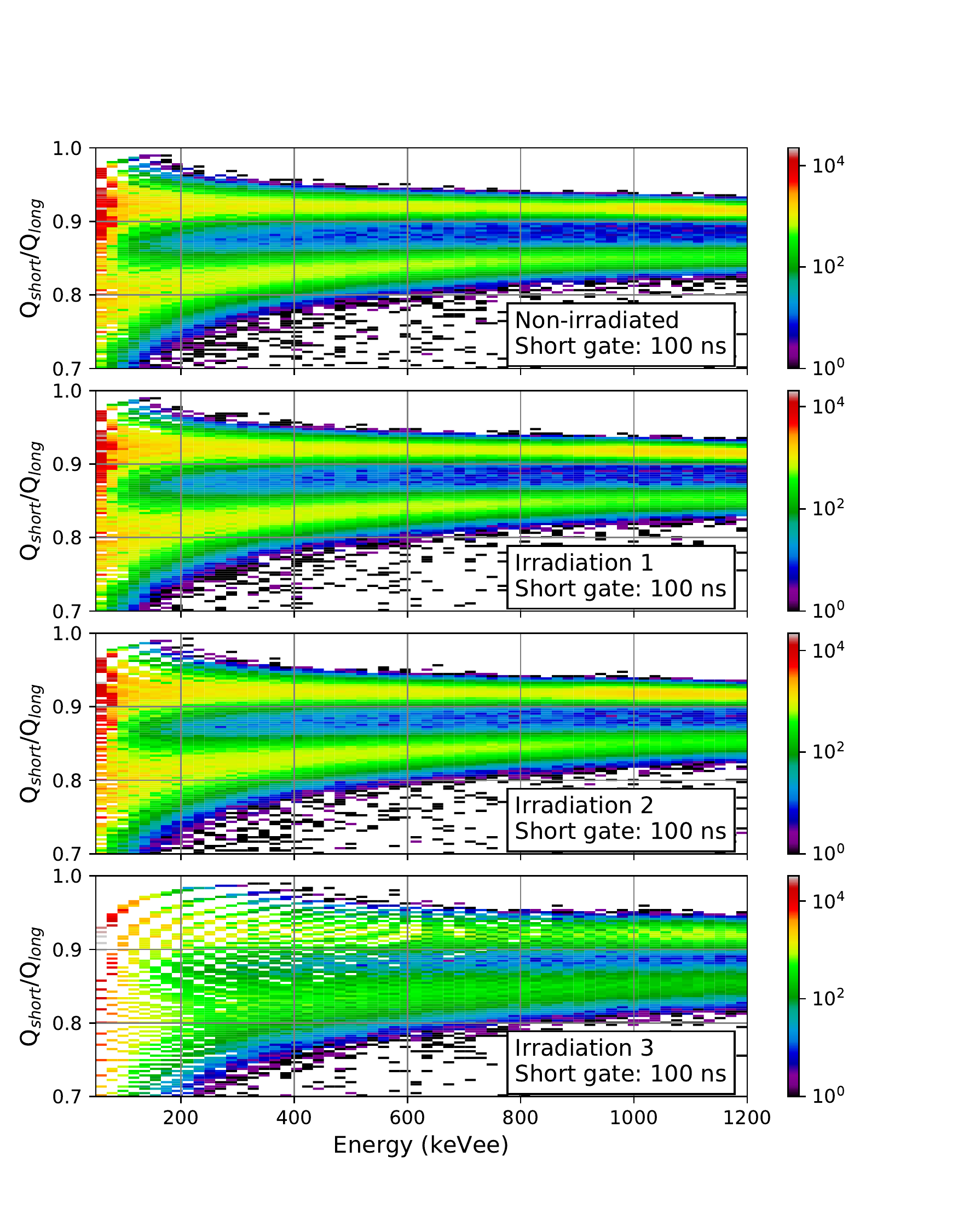}
  \caption{Example of PuBe n/$\gamma$ discrimination spectrum obtained for different irradiation time in the energy equivalent range up to 1200~keVee.}
  \label{fig:PSD_AllIrr}
 \end{figure}

For a better understanding of n/$\gamma$ discrimination, the length of the short gate was changed in 10~ns steps. The long gate was set to 250~ns and pre-gate to 50~ns (see Figure \ref{fig:pulse}). These values were constant for all measurements. It is easy to understand, that due to lower decay time in a pulse response of EJ-276 scintillator to gamma quanta, the charge integration value is higher for gamma rays than for neutrons ($Q_{short}^{\gamma}>Q_{short}^{n}$) in the same energy equivalent range. It corresponds to the higher ''banana-like'' branch spectrum part, like in Figure \ref{fig:PSD_Irr0}, which shows the shape changes in the plot for a different short gate and non-irradiated scintillator. On the other hand Figure \ref{fig:PSD_AllIrr} shows the discrimination spectra for the same short gate (100~ns) and the same energy range up to 1200~keVee but for different neutron irradiation time.

\begin{figure}[!htb]
  \centering
  \def\svgwidth{200pt}
  \includegraphics[width=1\textwidth]{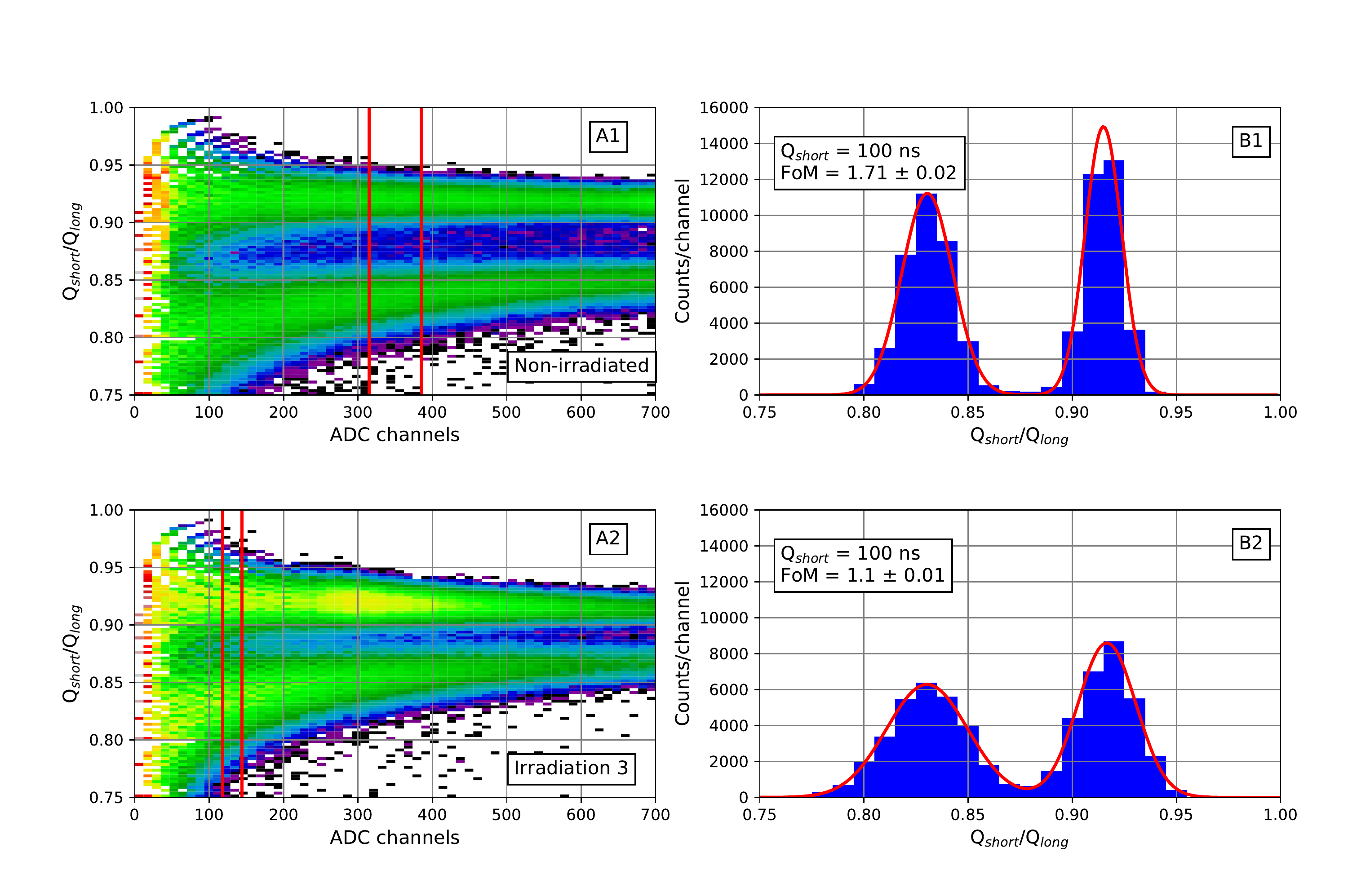}
  \caption{Example of n/$\gamma$ discrimination analysis for non-irradiated scintillator (A1) and with the longest irradiation time (A2). Short integration gate was set to 100~ns. B1 and B2 plots show the counts projection to Qshort /Qlong axis in the energy range of (500$\pm$50)~keVee. 
Red curves correspond to double Gaussian fit.}
  \label{fig:fom_irr}
 \end{figure}

For such conditions and the chosen energy range, the FoM defined by Equation \ref{eq:fom} was calculated:  
\begin{equation}
{FoM} = \frac{X_{n}-X_{\gamma}}{FWHM_n+FWHM_{\gamma}},
\label{eq:fom}
\end{equation}
where  X$_{n,\gamma}$ is the centroid position and FWHM$_{n,\gamma}$ is the width of the Gaussian distribution in a certain energy range. An example of analysis for a non-irradiated scintillator and with the longest irradiation time and energy range of (500$\pm$50)~keVee is presented in Figure \ref{fig:fom_irr}. In comparison to the longest irradiation times, it is seen that if we follow the chosen energy range for different irradiation time, the light yield reduction in the scintillator is responsible for the sensitivity change of our experimental set-up. With the same ADC charge sensitivity we are able to detect higher particle energy but with a strong reduction of particle discrimination. 

\begin{figure}[!htb]
  \centering
  \def\svgwidth{200pt}
  \includegraphics[width=1.1\textwidth]{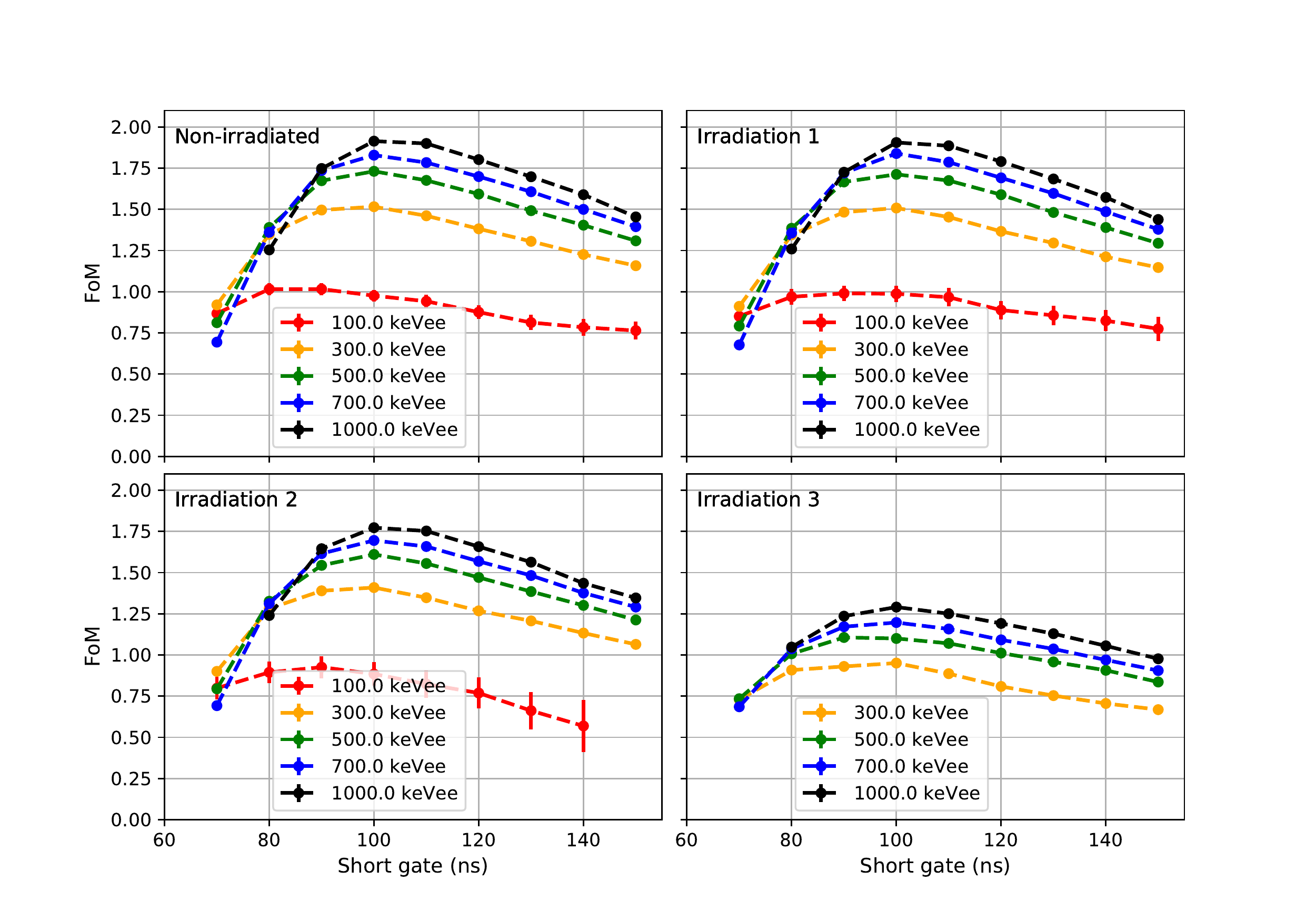}
  \caption{The figure of merit calculation as a dependence of short integration gate and different energy window.}
  \label{fig:fom_all}
 \end{figure}

\begin{figure}[!htb]
  \centering
  \def\svgwidth{200pt}
  \includegraphics[width=1.1\textwidth]{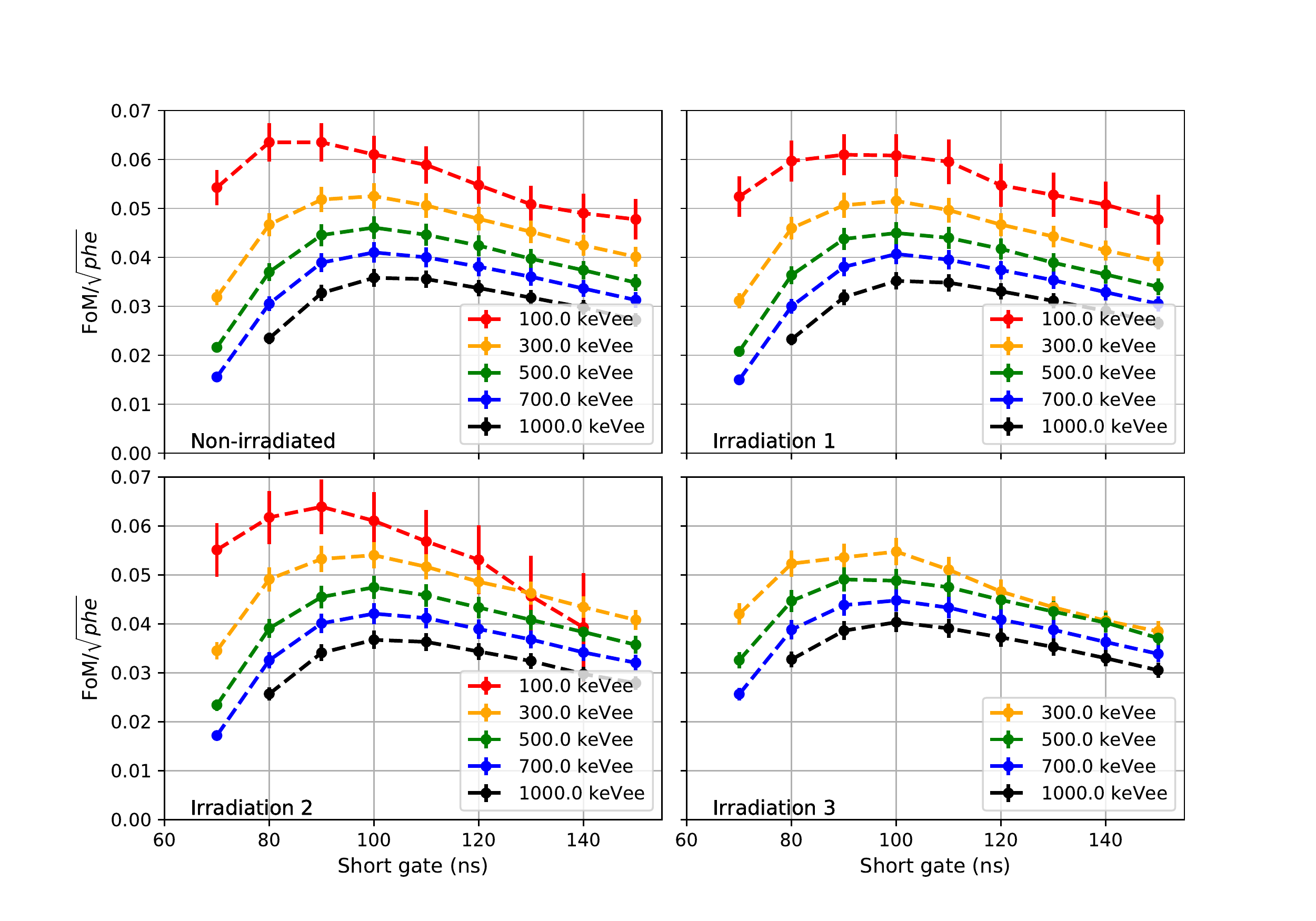}
  \caption{The figure of merit calculation normalised to a square root of the photo-electron number for each sample as a dependence of short integration gate and different energy window.}
  \label{fig:fom_all_norm}
 \end{figure}

Figure \ref{fig:fom_all} shows the obtained results of FoM calculation for all cases. The discrimination for a 100~keV energy range becomes impossible for the highest neutron fluence - there is no distinction between two Gaussian distributions. For a higher energy range, the n/$\gamma$ discrimination is still possible, but FoM reduction is in the range of 30$\%$. The last figure (Figure \ref{fig:fom_all_norm}) shows the same data but normalised to a square root of photo-electron number for a certain energy. It is worth noting, that for all four cases the values obtained are in good agreement. It suggests that only the light yield reduction is responsible for the FoM deterioration.

\section{Summary}
	The results of EJ-276 scintillator irradiation by fast neutrons were presented in this paper. It was shown that with neutron fluence increase the changes in emission and absorption spectrum are visible. The irradiation process expands the absorption spectrum which overlaps with the emission spectrum. As a consequence, light yield reduction is observed. Finally, we have shown that for irradiation with fluence up to 10$^{13}$n/cm$^2$ no significant changes in initial scintillator properties are observed. The neutron fluence in the range of 10$^{15}$~n/cm$^2$ seems to be the limitation in n/$\gamma$ discrimination, but the obtained value is still very high in most nuclear and high energy physics applications.  

\section*{References} 
\bibliography{iopart-num}

\end{document}